%% file: main.tex
\pdfoutput=1
\documentclass[12pt,a4paper]{article}

\usepackage{ifthen} 
\newboolean{pdflatex}
\setboolean{pdflatex}{true} 

\newboolean{articletitles}
\setboolean{articletitles}{true} 

\newboolean{uprightparticles}
\setboolean{uprightparticles}{false} 

\usepackage{tikz}
\usetikzlibrary{patterns,fadings,shapes.arrows,shadows,snakes,arrows}
\usepackage{pgfplots}
\usepgfplotslibrary{dateplot}
\pgfplotsset{compat=1.12}
\usepgfplotslibrary{fillbetween}
\usepackage{cancel}

\newboolean{wordcount}
\setboolean{wordcount}{false} 

\def\paperauthors{LHCb collaboration} 
\def\paperasciititle{Fraction of chi_c decays in prompt J/psi production measured in pPb collisions at sqrt{s_{NN}}=8.16 TeV} 
\def\papertitle{Fraction of \chic decays in prompt \jpsi production measured in \pPb collisions at $\sqsnn=8.16\tev$} 
\def\paperkeywords{{High Energy Physics}, {LHCb}, {Quarkonia}, {Heavy Ion}} 
\def\papercopyright{\the\year\ CERN for the benefit of the LHCb collaboration} 
\def\paperlicence{CC BY 4.0 licence}
\def\paperlicenceurl{https://creativecommons.org/licenses/by/4.0/}

\input{preamble}
\def\fchic {\ensuremath{F_{\chic\to\jpsi}}\xspace}
\def\ptjpsi{\ensuremath{p_{\mathrm{T},J/\psi}}\xspace}
\def\ptgamma{\ensuremath{p_{\mathrm{T},\gamma}}\xspace}
\def\pPb {$p$Pb\xspace}
\def\Pbp {Pb$p$\xspace}
\def\epos {\mbox{\textsc{EPOS}}\xspace}
\def\cgamma {\ensuremath{\bcancel{\gamma}}\xspace}

\usepackage{longtable} 

\begin{document}

\renewcommand{\thefootnote}{\fnsymbol{footnote}}
\setcounter{footnote}{1}

\ifthenelse{\boolean{wordcount}}{}{
\input{title-LHCb-PAPER}}

\renewcommand{\thefootnote}{\arabic{footnote}}
\setcounter{footnote}{0}

\cleardoublepage


\pagestyle{plain} 
\setcounter{page}{1}
\pagenumbering{arabic}


\input{body}

\ifthenelse{\boolean{wordcount}}{}{
\input{acknowledgements}}

\ifthenelse{\boolean{wordcount}}{}{%



\addcontentsline{toc}{section}{References}
\input{main.bbl}

\newpage
\input{Authorship_LHCb-PAPER-2023-028}
}

\end{document}

%% file: preamble.tex
\usepackage[top=1in, bottom=1.25in, left=1in, right=1in]{geometry}

\usepackage{microtype}
\usepackage{lineno}  
\usepackage{xspace} 
\usepackage{caption} 

\usepackage{graphicx}  
\usepackage{color}
\usepackage{colortbl}
\graphicspath{{.}} 

\usepackage{amsmath} 
\usepackage{amssymb}
\usepackage{amsfonts}
\usepackage{upgreek} 

\newcommand*\patchAmsMathEnvironmentForLineno[1]{%
\expandafter\let\csname old#1\expandafter\endcsname\csname #1\endcsname
\expandafter\let\csname oldend#1\expandafter\endcsname\csname
end#1\endcsname
 \renewenvironment{#1}%
   {\linenomath\csname old#1\endcsname}%
   {\csname oldend#1\endcsname\endlinenomath}%
}
\newcommand*\patchBothAmsMathEnvironmentsForLineno[1]{%
  \patchAmsMathEnvironmentForLineno{#1}%
  \patchAmsMathEnvironmentForLineno{#1*}%
}
\AtBeginDocument{%
\patchBothAmsMathEnvironmentsForLineno{equation}%
\patchBothAmsMathEnvironmentsForLineno{align}%
\patchBothAmsMathEnvironmentsForLineno{flalign}%
\patchBothAmsMathEnvironmentsForLineno{alignat}%
\patchBothAmsMathEnvironmentsForLineno{gather}%
\patchBothAmsMathEnvironmentsForLineno{multline}%
\patchBothAmsMathEnvironmentsForLineno{eqnarray}%
}

\usepackage{hyperxmp}

\usepackage[pdftex,
            pdfauthor={\paperauthors},
            pdftitle={\paperasciititle},
            pdfkeywords={\paperkeywords},
            pdfcopyright={Copyright (C) \papercopyright},
            pdflicenseurl={\paperlicenceurl}]{hyperref}

\usepackage[colorinlistoftodos,textsize=scriptsize]{todonotes}

\usepackage[bottom,flushmargin,hang,multiple]{footmisc}

\usepackage[all]{hypcap} 

\input{lhcb-symbols-def} 

\usepackage{cite} 
\usepackage{mciteplus}

%% file: lhcb-symbols-def.tex
\usepackage{xspace} 
\usepackage{upgreek}

\def\lhcb   {\mbox{LHCb}\xspace}

\def\cms    {\mbox{CMS}\xspace}

\def\lhc    {\mbox{LHC}\xspace}

\def\velo   {VELO\xspace}

\def\spd    {SPD\xspace}

\def\ecal   {ECAL\xspace}
\def\hcal   {HCAL\xspace}
\def\MagUp {\mbox{\em Mag\kern -0.05em Up}\xspace}

\ifthenelse{\boolean{uprightparticles}}%
{

 \def\Peta        {\ensuremath{\upeta}\xspace}

 \def\Pmu         {\ensuremath{\upmu}\xspace}

 \def\Ppi         {\ensuremath{\uppi}\xspace}

 \def\Pchi        {\ensuremath{\upchi}\xspace}                 
 \def\Ppsi        {\ensuremath{\uppsi}\xspace}

 \def\PDelta      {\ensuremath{\Delta}\xspace}                 
 \def\PXi         {\ensuremath{\Xi}\xspace}                 
 \def\PLambda     {\ensuremath{\Lambda}\xspace}                 
 \def\PSigma      {\ensuremath{\Sigma}\xspace}                 
 \def\POmega      {\ensuremath{\Omega}\xspace}                 
 \def\PUpsilon    {\ensuremath{\Upsilon}\xspace}
 \let\oldPi\Pi
 \def\PPi         {\ensuremath{\oldPi}\xspace}

 \def\PB      {\ensuremath{\mathrm{B}}\xspace}                 
                  
 \def\PD      {\ensuremath{\mathrm{D}}\xspace}

 \def\PJ      {\ensuremath{\mathrm{J}}\xspace}                 
 \def\PK      {\ensuremath{\mathrm{K}}\xspace}

 \def\Pb      {\ensuremath{\mathrm{b}}\xspace}                 
 \def\Pc      {\ensuremath{\mathrm{c}}\xspace}

 \def\Pi      {\ensuremath{\mathrm{i}}\xspace}

 \def\Pp      {\ensuremath{\mathrm{p}}\xspace}

 \def\Ps      {\ensuremath{\mathrm{s}}\xspace}

 \def\thebaroffset{0.0em}
}
{

 \def\Peta        {\ensuremath{\eta}\xspace}

 \def\Pmu         {\ensuremath{\mu}\xspace}

 \def\Ppi         {\ensuremath{\pi}\xspace}

 \def\Pchi        {\ensuremath{\chi}\xspace}                 
 \def\Ppsi        {\ensuremath{\psi}\xspace}                 
                  
 \mathchardef\PDelta="7101
 \mathchardef\PXi="7104
 \mathchardef\PLambda="7103
 \mathchardef\PSigma="7106
 \mathchardef\POmega="710A
 \mathchardef\PUpsilon="7107
 \mathchardef\PPi="7105
                  
 \def\PB      {\ensuremath{B}\xspace}                 
                  
 \def\PD      {\ensuremath{D}\xspace}

 \def\PJ      {\ensuremath{J}\xspace}                 
 \def\PK      {\ensuremath{K}\xspace}

 \def\Pb      {\ensuremath{b}\xspace}                 
 \def\Pc      {\ensuremath{c}\xspace}

 \def\Pi      {\ensuremath{i}\xspace}

 \def\Pp      {\ensuremath{p}\xspace}

 \def\Ps      {\ensuremath{s}\xspace}

 \def\thebaroffset{0.18em}
}
\newcommand{\offsetoverline}[2][\thebaroffset]{\kern #1\overline{\kern -#1 #2}}%

\makeatletter
\ifcase \@ptsize \relax
  \newcommand{\miniscule}{\@setfontsize\miniscule{4}{5}}
\or
  \newcommand{\miniscule}{\@setfontsize\miniscule{5}{6}}
\or
  \newcommand{\miniscule}{\@setfontsize\miniscule{5}{6}}
\fi
\makeatother

\DeclareRobustCommand{\optbar}[1]{\shortstack{{\miniscule (\rule[.5ex]{1.25em}{.18mm})}
  \\ [-.7ex] $#1$}}

\def\mup        {{\ensuremath{\Pmu^+}}\xspace}
\def\mun        {{\ensuremath{\Pmu^-}}\xspace} 

\def\squark    {{\ensuremath{\Ps}}\xspace}

\def\cquark    {{\ensuremath{\Pc}}\xspace}
\def\cquarkbar {{\ensuremath{\overline \cquark}}\xspace}
\def\ccbar     {{\ensuremath{\cquark\cquarkbar}}\xspace}
\def\bquark    {{\ensuremath{\Pb}}\xspace}
\def\bquarkbar {{\ensuremath{\overline \bquark}}\xspace}
\def\bbbar     {{\ensuremath{\bquark\bquarkbar}}\xspace}

\def\pion   {{\ensuremath{\Ppi}}\xspace}
\def\piz    {{\ensuremath{\pion^0}}\xspace}
\def\pip    {{\ensuremath{\pion^+}}\xspace}
\def\pim    {{\ensuremath{\pion^-}}\xspace}

\def\KorKbar {\kern \thebaroffset\optbar{\kern -\thebaroffset \PK}{}\xspace}

\newcommand{\etaz}{\ensuremath{\Peta}\xspace}

\def\D       {{\ensuremath{\PD}}\xspace}

\def\DorDbar {\kern \thebaroffset\optbar{\kern -\thebaroffset \PD}\xspace}
\def\Dz      {{\ensuremath{\D^0}}\xspace}

\def\Dp      {{\ensuremath{\D^+}}\xspace}
\def\Dm      {{\ensuremath{\D^-}}\xspace}

\def\DpDm    {\ensuremath{\Dp {\kern -0.16em \Dm}}\xspace}

\def\B       {{\ensuremath{\PB}}\xspace}

\def\BorBbar {\kern \thebaroffset\optbar{\kern -\thebaroffset \PB}\xspace}

\def\Bd      {{\ensuremath{\B^0}}\xspace}

\def\BdorBdbar {\kern \thebaroffset\optbar{\kern -\thebaroffset \Bd}\xspace}
\def\Bu      {{\ensuremath{\B^+}}\xspace}

\def\Bp      {{\ensuremath{\Bu}}\xspace}

\def\Bs      {{\ensuremath{\B^0_\squark}}\xspace}

\def\BsorBsbar {\kern \thebaroffset\optbar{\kern -\thebaroffset \Bs}\xspace}

\def\jpsi     {{\ensuremath{{\PJ\mskip -3mu/\mskip -2mu\Ppsi}}}\xspace}
\def\psitwos  {{\ensuremath{\Ppsi{(2S)}}}\xspace}

\def\chic     {{\ensuremath{\Pchi_\cquark}}\xspace}
\def\chiczero {{\ensuremath{\Pchi_{\cquark 0}}}\xspace}
\def\chicone  {{\ensuremath{\Pchi_{\cquark 1}}}\xspace}
\def\chictwo  {{\ensuremath{\Pchi_{\cquark 2}}}\xspace}

\def\Upsilonres  {{\ensuremath{\PUpsilon}}\xspace}
\def\Y#1S{\ensuremath{\PUpsilon{(#1S)}}\xspace}

\def\proton      {{\ensuremath{\Pp}}\xspace}

\def\LorLbar     {\kern \thebaroffset\optbar{\kern -\thebaroffset \PLambda}\xspace}

\def\BF         {{\ensuremath{\mathcal{B}}}\xspace}
\def\BR         {\BF}

\def\to                 {\ensuremath{\rightarrow}\xspace}

\def\AT#1     {\ensuremath{A_{\mathrm{T}}^{#1}}\xspace}           

\def\C#1      {\ensuremath{\mathcal{C}_{#1}}\xspace}                       
\def\Cp#1     {\ensuremath{\mathcal{C}_{#1}^{'}}\xspace}                    
\def\Ceff#1   {\ensuremath{\mathcal{C}_{#1}^{\mathrm{(eff)}}}\xspace}        
\def\Cpeff#1  {\ensuremath{\mathcal{C}_{#1}^{'\mathrm{(eff)}}}\xspace}       
\def\Ope#1    {\ensuremath{\mathcal{O}_{#1}}\xspace}                       
\def\Opep#1   {\ensuremath{\mathcal{O}_{#1}^{'}}\xspace}                    

\newcommand{\aunit}[1]{\ensuremath{\text{\,#1}}}       

\newcommand{\tev}{\aunit{Te\kern -0.1em V}\xspace}
\newcommand{\gev}{\aunit{Ge\kern -0.1em V}\xspace}
\newcommand{\mev}{\aunit{Me\kern -0.1em V}\xspace}
\newcommand{\kev}{\aunit{ke\kern -0.1em V}\xspace}
\newcommand{\ev}{\aunit{e\kern -0.1em V}\xspace}
 
\newcommand{\mevc}{\ensuremath{\aunit{Me\kern -0.1em V\!/}c}\xspace}
\newcommand{\gevc}{\ensuremath{\aunit{Ge\kern -0.1em V\!/}c}\xspace}
\newcommand{\mevcc}{\ensuremath{\aunit{Me\kern -0.1em V\!/}c^2}\xspace}
\newcommand{\gevcc}{\ensuremath{\aunit{Ge\kern -0.1em V\!/}c^2}\xspace}

\def\gsim{{~\raise.15em\hbox{$>$}\kern-.85em
          \lower.35em\hbox{$\sim$}~}\xspace}
\def\lsim{{~\raise.15em\hbox{$<$}\kern-.85em
          \lower.35em\hbox{$\sim$}~}\xspace}

\def\sqs   {\ensuremath{\protect\sqrt{s}}\xspace}
\def\sqsnn {\ensuremath{\protect\sqrt{s_{\scriptscriptstyle\text{NN}}}}\xspace}
\def\pt         {\ensuremath{p_{\mathrm{T}}}\xspace}

\def\evtgen     {\mbox{\textsc{EvtGen}}\xspace}

\def\geant      {\mbox{\textsc{Geant4}}\xspace}

\def\photos     {\mbox{\textsc{Photos}}\xspace}

\def\pythia     {\mbox{\textsc{Pythia}}\xspace}

\def\tell1  {TELL1\xspace}
\def\ukl1   {UKL1\xspace}

\newcommand{\lhcborcid}[1]{\href{https://orcid.org/#1}{\hspace*{0.1em}\raisebox{-0.45ex}{\includegraphics[width=1em]{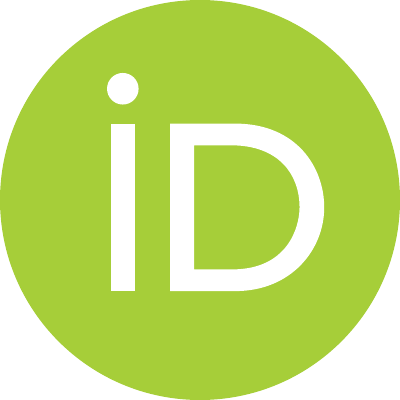}}}}


%% file: title-LHCb-PAPER.tex

\begin{titlepage}
\pagenumbering{roman}

\vspace*{-1.5cm}
\centerline{\large EUROPEAN ORGANIZATION FOR NUCLEAR RESEARCH (CERN)}
\vspace*{1.5cm}
\noindent
\begin{tabular*}{\linewidth}{lc@{\extracolsep{\fill}}r@{\extracolsep{0pt}}}
\ifthenelse{\boolean{pdflatex}}
{\vspace*{-1.5cm}\mbox{\!\!\!\includegraphics[width=.14\textwidth]{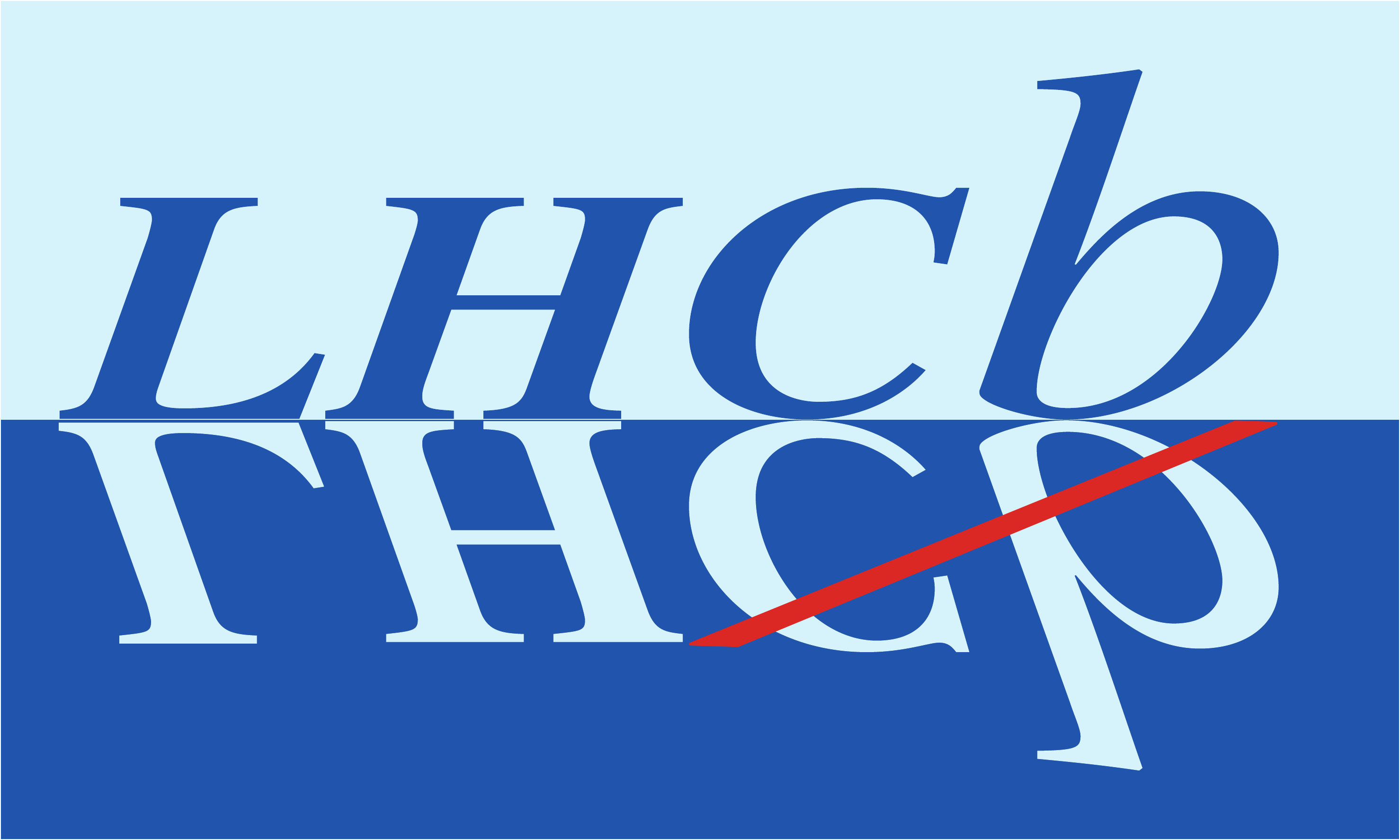}} & &}%
{\vspace*{-1.2cm}\mbox{\!\!\!\includegraphics[width=.12\textwidth]{lhcb-logo.eps}} & &}%
\\
 & & CERN-EP-2023-224 \\  
 & & LHCb-PAPER-2023-028 \\  
 & & November 02 2023 \\ 
 & & \\
\end{tabular*}

\vspace*{4.0cm}

{\normalfont\bfseries\boldmath\huge
\begin{center}
  \papertitle 
\end{center}
}

\vspace*{1.0cm}

\begin{center}
\paperauthors\footnote{Authors are listed at the end of this paper.}
\end{center}

\vspace{\fill}

\begin{abstract}
    \noindent The fraction of \chicone and \chictwo decays in the prompt \jpsi yield, \mbox{\fchic$=\sigma_{\chic\to\jpsi}/\sigma_{\jpsi}$}, is measured by the \lhcb detector in \pPb collisions at \mbox{$\sqsnn=8.16\tev$}. The study covers the forward \mbox{($1.5<y^*<4.0$)} and backward \mbox{($-5.0<y^*<-2.5$)} rapidity regions, where $y^*$ is the \jpsi rapidity in the nucleon-nucleon center-of-mass system. Forward and backward rapidity samples correspond to integrated luminosities of \mbox{13.6 $\pm$ 0.3 nb$^{-1}$} and \mbox{20.8 $\pm$ 0.5 nb$^{-1}$}, respectively. The result is presented as a function of the \jpsi transverse momentum \ptjpsi in the range \mbox{1$<\ptjpsi<20\gevc$}. The \fchic fraction at forward rapidity is compatible with the \lhcb measurement performed in $pp$ collisions at \mbox{$\sqs=7\tev$}, whereas the result at backward rapidity is \mbox{2.4 $\sigma$} larger than in the forward region for \mbox{$1<\ptjpsi<3\gevc$}. The increase of \fchic at low \ptjpsi at backward rapidity is compatible with the suppression of the \psitwos contribution to the prompt \jpsi yield. The lack of in-medium dissociation of \chic states observed in this study sets an upper limit of 180\mev on the free energy available in these \pPb collisions to dissociate or inhibit charmonium state formation.
\end{abstract}

\vspace*{1.0cm}

\begin{center}
  Submitted to
  Phys.~Rev.~Lett. 
\end{center}

\vspace{\fill}

{\footnotesize 
\centerline{\copyright~\papercopyright. \href{\paperlicenceurl}{\paperlicence}.}}
\vspace*{2mm}

\end{titlepage}


\newpage
\setcounter{page}{2}
\mbox{~}
%
%
%
%

%% file: body.tex
\input{Introduction}

\input{detector}

\input{data_selection}

\input{signal_couting}

\input{systematic_error}

\input{result}

%% file: Introduction.tex
Heavy ion collisions in the relativistic regime provide an opportunity to release quarks and gluons from hadrons and form a hot and dense quark-gluon plasma (QGP), the same state of matter theorized to exist microseconds after the Big Bang and to compose the core of neutron stars \cite{Annala:2019puf}. While there are many indications of QGP formation in nucleus-nucleus collisions at RHIC and the \lhc, its formation in small systems such as proton-nucleus collisions is not firmly established. The observation of collective particle flow in \pPb collisions at the LHC, and $p$A and $d$A collisions at RHIC, suggests the existence of QGP droplets in these collisions (see for example \cite{PHENIX:2018lia}), but a lack of other signatures prevents a conclusion.
Quarkonium states have a broad range of binding energies \cite{Satz:2005hx} of the same order of magnitude as the freeze-out temperature, the minimum temperature needed to form a QGP \cite{HotQCD:2018pds,Sharma:2018jqf,Flor:2021olm}. The spectrum of quarkonium states that survive nucleus collisions is a powerful tool to determine the free energy and temperature reached by the initial stage of heavy ion collisions. However, a quarkonium state can also be broken by its interaction with comoving particles if the particle multiplicity is high enough, as observed in the nucleus-going direction in proton- and deuteron-nucleus collisions \cite{Ferreiro:2014bia}. 
The nuclear modification factor, defined as the ratio between the particle yield per nucleon interaction measured in heavy ion and $pp$ collisions, is the most commonly used observable to quantify nuclear effects.
The spin-1 ground-state charmonium meson \jpsi, with a binding energy of 640\mev, has a nuclear modification factor similar to the open charm meson \Dz at forward and backward rapidity ranges in \pPb collisions~\cite{LHCb-PAPER-2017-015}. This observation  indicates that the nuclear modification factor observed in \jpsi yields in these collisions can be attributed solely to nuclear modification of parton densities before the formation of the \ccbar pair. 
The charmonium state \psitwos, with a binding energy of 50\mev, has a suppression stronger than the \jpsi state at backward rapidity in $p(d)$A collisions at RHIC \cite{PHENIX:2022nrm} and the LHC \cite{LHCb-PAPER-2015-058,ALICE:2014cgk,LHCb-PAPER-2023-024}, indicating that this weakly bound state is also suppressed by final-state nuclear effects.

Additional constraints on the free local energy produced in $p$A collisions, which would dissociate or prevent charmonium state formation, can be provided by the \chicone and \chictwo states. These have binding energies of 220 and $180\mev$, respectively.
The P-wave charmonium states $\chi_{cn}$ are mainly reconstructed from the radiative decay $\chi_{cn}\to\jpsi\gamma$ with branching ratios 1.4\%, 34\% and 19\% for $n=0,1$ and $2$ \cite{PDG2022}. 
The production of \chiczero mesons is rarely studied in this decay mode, due to its small branching ratio.
Measurements of \chic production typically require high detection efficiency for low-energy photons, and effective discrimination against the overwhelming \piz decay sources producing large combinatorial backgrounds in the $\jpsi\gamma$ mass distribution. 
For these reasons, \chic measurements in heavy ion collisions are rare. There are two \chic measurements in nucleus collisions: (i) HERA-B measured the fraction of \chic decays in \jpsi production in proton on C and W targets at \mbox{$\sqsnn=41.6$\gev} \cite{HERA-B:2008rhh}, showing no dependence on the target material and (ii) PHENIX measured the \chic/\jpsi fraction in $d$Au collisions at \mbox{$\sqsnn=200$\gev} \cite{PHENIX:2013pmn}, consistent with measurements in $pp$ collisions at the same energy within the large statistical uncertainty.
The identified \chicone and \chictwo yields measured by the \lhcb collaboration in \pPb collisions at \mbox{$\sqsnn=8.16$\tev} are consistent \cite{LHCb-PAPER-2020-048}, although with large uncertainties given the difficulty in resolving the mass peaks. 

This Letter reports, for the first time at the \lhc, the fraction \fchic of \mbox{$\chic\to(\jpsi\to\mup\mun)\gamma$} decays in prompt \mbox{\jpsi\to\mup\mun} yields. The measurement is made in eight \ptjpsi ranges over \mbox{$1<\ptjpsi<20$\gevc} in \pPb collisions at \mbox{$\sqsnn=8.16$\tev} at forward \mbox{($1.5<y^*<4.0$)} and backward \mbox{($-5.0<y^*<-2.5$)} rapidities, where $y^*$ is the \jpsi rapidity in the nucleon-nucleon center-of-mass system. 
Prompt \jpsi mesons are produced directly in the hadronization process (direct contribution) or via decays of higher-mass charmonium states (feeddown contribution).
The measured \chic yield is the sum of the \chicone and \chictwo yields, which minimizes the uncertainties related to the separation of these states. 
The fraction \fchic is obtained by
\begin{equation}
    \label{eq:f_chic}
    \fchic \equiv\frac{\sigma_{\chic\to\jpsi\gamma}}{\sigma_{\jpsi}}=\frac{N_{\chic\to\jpsi\gamma}}{N_{\jpsi}~\varepsilon_{\chic/\jpsi}},
\end{equation}
\noindent where $N_{\chic\to\jpsi\gamma}$ and $N_{\jpsi}$ are the prompt \chic\to\jpsi$\gamma$ and \jpsi yields, and $\varepsilon_{\chic/\jpsi}$ is the fraction of \chic\to\jpsi$\gamma$ decays which are detected in the \lhcb acceptance relative to the decays where only the \jpsi is detected.

%% file: detector.tex
The LHCb detector is a single-arm forward spectrometer described in Refs. \cite{LHCb-DP-2008-001,LHCb-DP-2014-002}. The silicon-strip vertex detector (\velo) surrounding the interaction region allows the determination of the position of the collision point, the primary vertex (PV). Charged particle tracks are determined by the combination of hits in the \velo, a large-area silicon-strip detector located upstream of a dipole magnet with a bending power of about 4 Tm, and three stations of silicon-strip detectors and straw drift tubes (OT) placed downstream of the magnet. Muons are identified by a system composed of alternating layers of iron and multiwire proportional chambers behind electromagnetic and hadronic calorimeters. Photons are identified by the calorimeter system \cite{LHCb-PROC-2014-017} consisting of a scintillating pad detector (\spd), a preshower system (PS), an electromagnetic (\ecal) calorimeter, and a hadronic (\hcal) calorimeter. The \spd and PS are designed to discriminate between signals from photons and electrons, while the \ecal and \hcal provide the energy measurement and identification of photons and neutral hadrons. The relative energy resolution of the \ecal is \mbox{8\%/$\sqrt{E}+0.9$\%}, where $E$ is in \gev.

This analysis is based on data acquired during the 2016 \lhc heavy-ion run, where the \lhcb experiment recorded proton and $^{208}$Pb ion collisions at a center-of-mass energy per nucleon pair of \mbox{\sqsnn = 8.16 \tev}. 
The forward (positive) rapidity sample is collected in collisions where the proton follows the direction from the \velo to the muon detectors, namely \pPb collisions, corresponding to an integrated luminosity of \mbox{$13.6\pm0.3$ nb$^{-1}$}. 
The backward (negative) rapidity sample is obtained with a reverse beam direction, \Pbp collisions, with \mbox{$20.8\pm0.5$ nb$^{-1}$} integrated luminosity.

%% file: data_selection.tex
The online event selection used in this analysis is performed by a trigger system consisting of a hardware stage that selects events containing at least one muon candidate, and two software trigger stages in which events with two tracks identified as muons with \mbox{$\pt>500$\mevc} are selected. The muon pair is required to have an invariant mass within 150 \mevcc of the known \jpsi mass \cite{PDG2022}.
In the offline selection, muons are identified by a neural network algorithm. The muon also must satisfy the momentum requirements \mbox{$\pt>600$\mevc}, \mbox{$p>8$\gevc}, and be in the \lhcb pseudorapidity range \mbox{$2<\eta<5$}. The \jpsi candidate is selected by requiring the muon pair invariant mass to be within 80 \mevcc of its known mass and its transverse momentum \ptjpsi must be larger than 1\gevc. The \jpsi candidate must be consistent with originating from the collision point.

Photons are identified as isolated clusters in the \ecal. The cluster must not belong to a charged particle or a \piz decay with 80\% confidence level as determined by neural network algorithms. The photon must be in the \ecal acceptance \mbox{$2<\eta<4.5$} and have transverse momentum $\ptgamma>400$\mevc.

The detector performance for the \jpsi and \chic signals are studied using simulated samples generated by \pythia \cite{Sjostrand:2007gs} and embedded with events generated by \epos \cite{Pierog:2013ria}, which accounts for the underlying event activity of \pPb collisions. Decays of short-lived particles are performed with the \evtgen decay package \cite{Lange:2001uf}. Radiative QED corrections to the decays containing charged particles in the final state are applied with the \photos package \cite{Golonka:2005pn}. The generated \chicone and \chictwo decays are required to have both muons from the \jpsi decay in the \lhcb acceptance. The response of the \lhcb detector is modelled using \geant \cite{LHCb-PROC-2011-006}. 
Weights are assigned to the simulated events such that the \velo cluster multiplicity matches that of the data. The \lhcb detection efficiency correction $\varepsilon_{\chic/\jpsi}$ is obtained from the simulated \chic sample using \mbox{Eq.~(\ref{eq:f_chic})}, where $\fchic=1$ by definition.

%% file: signal_couting.tex
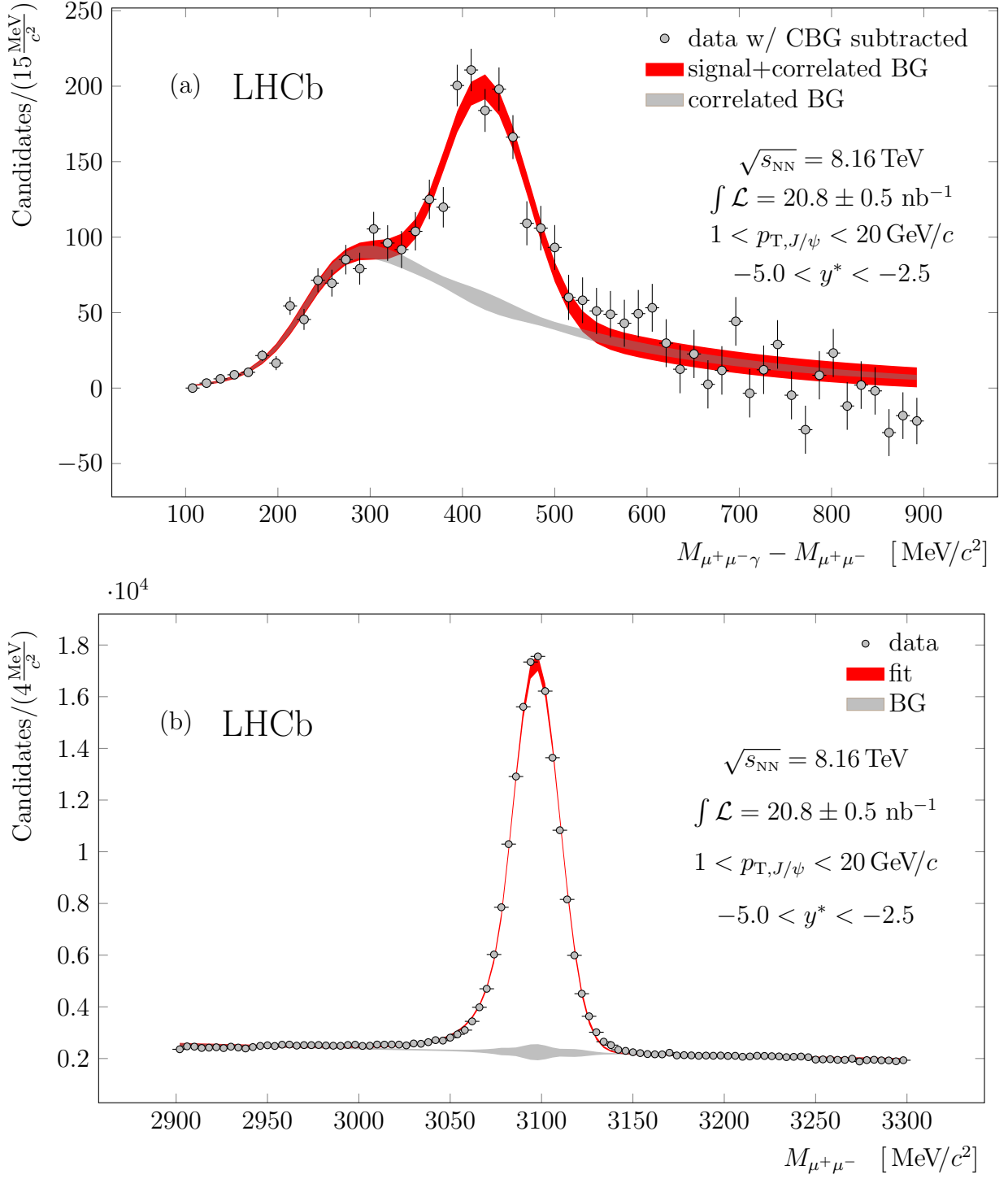
\begin{figure}[!htb]
    \nolinenumbers
    \input{Fig1}
    \caption{(a) Difference in the invariant mass of \mup\mun$\gamma$ and \mup\mun combinations in the \chic region after combinatorial background subtraction. The red and gray bands represent the fit result and the physical background components, respectively. The widths of the bands represent 68\% CL. (b) Invariant mass distribution of \mup\mun pairs in the \jpsi mass region along with the fitted function.}
    \label{fig:chic_peaks}
\end{figure}

The selected sample is overwhelmingly composed of combinatorial background (CBG). The remaining components are correlated background, also called physics background, and the signal \chic decays. 
This analysis uses the invariant mass difference \mbox{$\Delta M=M_{\mup\mun\gamma}-M_{\mup\mun}$} for fitting, minimizing the impact of the muon pair mass resolution.  
The shape of the combinatorial background distribution, $Y_{\rm CBG}(\Delta M)$, is determined by mixing \jpsi decays and photons from different events. 
The two events must have similar track multiplicities and collision vertex positions.
The overall combinatorial background yield $N_{\rm CBG}$ is determined by normalizing the CBG shape $Y_{\rm CBG}$ to have the same integral as the default sample in the mass region $700<\Delta M<900\mevcc$, where the signal and the correlated background are expected to be negligible. Figure \ref{fig:chic_peaks}(a) shows the $\Delta M$ distribution after the combinatorial background subtraction. 

The correlated background is composed of radiative \jpsi decays (\mbox{$\jpsi\to \mup\mun\gamma$}) and  partially reconstructed \mbox{$\psitwos\to\jpsi\piz\piz\to\mup\mun\gamma$} decays which are studied using simulated samples. The partially reconstructed \psitwos contribution is 8 (30) times smaller than the \jpsi radiative decay at high (low) \ptjpsi when considering its measured yield  relative to \jpsi decays~\cite{LHCb-PAPER-2023-024} and the branching ratio $\BR(\psitwos\to\jpsi\piz\piz)$ \cite{PDG2022}. The $\Delta M$ distribution accounting for the total correlated background is described by
\begin{equation}
    \label{eq:corr}
    Y_{\rm corr}(\Delta M) = \frac{A_{\rm corr} e^{B\cdot \Delta M}}{1+e^{-\frac{\Delta M - \Delta M_0}{\sigma_{\Delta M_b}}}},
\end{equation}
\noindent with \ptgamma-dependent parameters $B$, $\Delta M_0$ and $\sigma_{\Delta M_b}$ initially determined from the simulation and $A_{\rm corr}$ is its normalization. The two \chic states are described by the sum of two Gaussian functions, $G$, with a common resolution term as
\begin{eqnarray}
        Y_{\chic}(\Delta M) &=& f_{\chicone}G(\Delta M; \Delta M_{\chicone},\sigma_{\Delta M}) \\\nonumber
        &+& (1-f_{\chicone})G(\Delta M; \Delta M_{\chicone}+\Delta M_{\rm 1,2},\sigma_{\Delta M}), \nonumber
\end{eqnarray}
\noindent where $0.4<f_{\chicone}<0.6$ is the contribution of the \chicone decay to the total \chic yield as obtained in Ref. \cite{LHCb-PAPER-2020-048}, \mbox{$\Delta M_{\chicone}=M_{\chicone}-M_{\jpsi}$} is the free parameter accounting for the mass difference of the \chicone state as measured by the detector, \mbox{$\Delta M_{\rm 1,2}=45.5$\mevcc} is the known mass difference between the \chicone and \chictwo states~\cite{PDG2022}. The total shape of the $\Delta M$ distribution accounting for the \chic signal and background is given by
\begin{eqnarray}
    \label{eq:fitting_function}
    Y_{\mup\mun\gamma}(\Delta M) &=& N_{\rm CBG} Y_{\rm CBG}(\Delta M) + N_{\rm corr} Y_{\rm corr}(\Delta M) \\\nonumber
    &+& N_{\chic\to\jpsi\gamma}Y_{\chic}(\Delta M),
\end{eqnarray}
\noindent where $N_{\rm corr}$ is the yield of the correlated background. 
The signal yields are obtained with a maximum-likelihood fit of Eq. (\ref{eq:fitting_function}) to the $\Delta M$ distributions, with the initial parameters determined from simulation.
The total \chic yields, integrated over \ptjpsi, are \mbox{$(11.8 \pm 0.6) \times 10^{3}$} and \mbox{$(15.2 \pm 0.8) \times 10^{3}$} in the forward and backward rapidity samples, respectively. The \chic yields are independently obtained in eight \ptjpsi ranges and are stable with Gaussian variation of the initial parameters determined from simulation. Figure \ref{fig:chic_peaks}(a) shows the $\Delta M$ distribution, integrated over \ptjpsi, with the fit results overlaid. 

The \jpsi yield is determined in eight \ptjpsi ranges using a maximum-likelihood fit to the $\mu\mu$ invariant mass distribution. The fitting function is defined by the sum of a Crystal Ball function, CB$_{\rm pdf}$, \cite{Skwarnicki:1986xj} for signal and an exponential function for background 
\begin{equation}
    Y_{\mu^+\mu^-}(M) = A e^{b M} + N_{\jpsi}{\rm CB}_{\rm pdf}(M; M_{\jpsi}, \sigma_M, \alpha, n)
    \label{eq:jpsi_fit}
\end{equation}
\noindent to the \mup\mun invariant mass distribution using the log-likelihood method. The parameters $A$ and $b$ define the scale and slope of the exponential components. The parameters $\alpha$ and $n$ are determined from a \ptjpsi-integrated fit and fixed for the fits in different \ptjpsi ranges. The parameter $M_{\jpsi}$ is fixed to the known value of the \jpsi mass~\cite{PDG2022}. The \ptjpsi-integrated $M_{\mup\mun}$ distribution and the results of the fit are shown in Fig.~\ref{fig:chic_peaks}(b).

%% file: Fig1.tex
\pgfplotsset{
    discard if not/.style 2 args={
        x filter/.code={
            \edef\tempa{\thisrow{#1}}
            \edef\tempb{#2}
            \ifx\tempa\tempb
            \else
                \def\pgfmathresult{inf}
            \fi
        }
    }
}
\ifthenelse{\boolean{wordcount}}{}{%
\begin{tikzpicture}
\begin{axis}[
legend columns=1,
legend style={draw=none, legend cell align={left}},
legend entries={data w/ CBG subtracted, signal+correlated BG, correlated BG},
width=1.0\linewidth,
height=0.6\linewidth,
ylabel={\hspace{4.5cm}Candidates/$(15 \frac{\rm MeV}{c^2})$},
xlabel={\hspace{9cm}$M_{\mup\mun\gamma}-M_{\mup\mun} ~~ \left[\mevcc\right]$}] 
\addplot[only marks, black, mark options={black, mark=*, fill=gray!50!white, scale=1.0}, discard if not={system}{Pbp}, 
         error bars/.cd, x dir=both, x explicit, y dir=both, y explicit, error mark=none] table  [x=M, y=data, x error expr=\thisrow{dM}/2, y error=data_err] {chic_inv_mass.data};
\addplot [name path=upper,draw=none,forget plot] table[x=M, y expr=\thisrow{fit}+\thisrow{fit_err}] {chic_inv_mass.data};
\addplot [name path=lower,draw=none,forget plot] table[x=M, y expr=\thisrow{fit}-\thisrow{fit_err}] {chic_inv_mass.data};
\addplot +[fill=red, opacity=1] fill between[of=lower and upper];
\addplot [name path=upper,draw=none,forget plot] table[x=M, y expr=\thisrow{bremm}+\thisrow{bremm_err}] {chic_inv_mass.data};
\addplot [name path=lower,draw=none,forget plot] table[x=M, y expr=\thisrow{bremm}-\thisrow{bremm_err}] {chic_inv_mass.data};
\addplot +[fill=gray, opacity=0.5] fill between[of=lower and upper];
\node at (axis cs: 100, 200) {(a)};
\node at (axis cs: 200, 200) {\Large LHCb};
\node at (axis cs: 800, 150) {$\sqsnn=8.16\tev$};
\node at (axis cs: 800, 125) {$\int \mathcal{L}=20.8\pm0.5$ nb$^{-1}$};
\node at (axis cs: 800, 100) {$1<\ptjpsi<20\gevc$};
\node at (axis cs: 800, 75) {$-5.0<y^*<-2.5$};
\end{axis}
\end{tikzpicture}
\begin{tikzpicture}
\begin{axis}[
legend entries={data~~~,fit~~~,BG},
legend style={draw=none, legend cell align={left}},
ylabel={\hspace{4cm}Candidates/$(4 \frac{\rm MeV}{c^2})$},
xlabel={\hspace{11cm}$M_{\mup\mun} ~~ \left[\mevcc \right]$},
width=1.0\linewidth,
height=0.6\linewidth,
/pgf/number format/.cd, 1000 sep={}]
\addplot[only marks, black, mark options={black, mark=*, fill=gray!50!white, scale=0.8},
         error bars/.cd, x dir=both, x explicit, y dir=both, y explicit, error mark=none] table  [x=M, y=data, x error =dM, y error=data_err] {jpsi_inv_mass.data};
\addplot [name path=upper,draw=none,forget plot] table[x=M, y expr=\thisrow{fit}+10*\thisrow{fit_err}] {jpsi_inv_mass.data};
\addplot [name path=lower,draw=none,forget plot] table[x=M, y expr=\thisrow{fit}-10*\thisrow{fit_err}] {jpsi_inv_mass.data};
\addplot +[fill=red, opacity=1] fill between[of=lower and upper];
\addplot [name path=upper,draw=none,forget plot] table[x=M, y expr=\thisrow{corr}+10*\thisrow{corr_err}] {jpsi_inv_mass.data};
\addplot [name path=lower,draw=none,forget plot] table[x=M, y expr=\thisrow{corr}-10*\thisrow{corr_err}] {jpsi_inv_mass.data};
\addplot +[fill=gray, opacity=0.5] fill between[of=lower and upper];
\node at (axis cs: 2900, 15000) {(b)};
\node at (axis cs: 2950, 1.5e4) {\Large LHCb};
\node at (axis cs: 3250, 1.35e4) {$\sqsnn=8.16\tev$};
\node at (axis cs: 3250, 1.15e4) {$\int \mathcal{L}=20.8\pm0.5$ nb$^{-1}$};
\node at (axis cs: 3250, 0.95e4) {$1<\ptjpsi<20\gevc$};
\node at (axis cs: 3250, 0.75e4) {$-5.0<y^*<-2.5$};
\end{axis}
\end{tikzpicture}\\
}

%% file: systematic_error.tex
Variation of the initial parameters when fitting the $\Delta M$ distribution and the fixed mass resolution parameter $\Delta\sigma_M$ cause the largest systematic uncertainties on the \chic yields, mostly for the lowest \ptjpsi interval. These variations account for potential multiple local minima in the log-likelihood function used in the fits. Any deviation from the Gaussian shape assumption for the \chic peaks is tested by comparing the yields obtained by fitting Eq.~(\ref{eq:fitting_function}) and from the integral over \mbox{$300 < \Delta M < 600$ \mevcc} after subtracting the fitted background contributions $Y_{\rm CBG}+Y_{\rm corr}$. The difference between the yield obtained from the fit and the integral is assigned as a systematic uncertainty on the yield. 
The statistical uncertainties associated to the mixed \jpsi and $\gamma$ event samples are negligible. However, variations on the mass range used to normalize the mixed event distribution $Y_{\rm CBG}$ contribute to the uncertainty in the \chic yields. The only significant uncertainty in the \jpsi yield determination comes from the difference between the result obtained from the mass peak fitting and the integral over the \jpsi peak region after subtracting the background component.

The photon detection efficiency is the main contribution to the factor $\varepsilon_{\chic/\jpsi}$.
A validation of the photon detection efficiency $\varepsilon_{\gamma}$ is performed by studying partially reconstructed \etaz decays in data and simulation. Clear peaks are observed in the \mbox{$\pip\pim\gamma$} invariant mass distribution (well separated from the fully reconstructed \mbox{$\eta\to\pip\pim\gamma$} peak) and the \mbox{$\pip\pim$} mass distribution coming from \mbox{$\etaz\to\pip\pim(\piz\to\gamma\cgamma$)} and \mbox{$\etaz\to\pip\pim(\piz\to\cgamma\cgamma)$} decays, where \cgamma is a missing photon. The detection efficiency in $\etaz$ decays is measured in data and simulation by 
\begin{equation}
    \varepsilon_{\gamma}(\ptgamma) = \frac{N_{\etaz\to\pip\pim(\piz\to\gamma\cgamma)}(\ptgamma)}{\mathcal{M}_{\gamma(\pi\pi)}N_{\etaz\to\pip\pim(\piz\to\cgamma\cgamma)}(p_{\rm T,\pip\pim})},
    \label{eq:photon_eff_measurement}
\end{equation}

\noindent where $\mathcal{M}_{\gamma(\pi\pi)}$ is a matrix, obtained from simulation, that unfolds the \ptgamma distribution from \mbox{$\etaz\to \pip\pim(\piz\to\cgamma\cgamma)$} decays as a function of \mbox{$p_{\rm T,\pip\pim}$}. The efficiencies measured in data and simulation are consistent within 3\% in the photon \pt range \mbox{$400<\ptgamma<5000\mevc$}. Initial-state effects in the nucleus on the input kinematics are accounted for in simulation by weighting events according to the nuclear parton density function EPPS21 \cite{Eskola:2021nhw}.

All the measurements assume \chic production with zero polarization. The effects of a potential \chic polarization on the detector efficiency are studied by weighting the simulated events according to different polarization scenarios as reported in Ref. \cite{LHCb-PAPER-2011-030}. The measured \jpsi polarization by \lhcb \cite{LHCb-PAPER-2013-008} and the observation that either \chicone or \chictwo states is strongly polarized by the \cms collaboration \cite{CMS:2019jas} poses constraints to the scenarios adopted when weighting simulated events. The standard deviation of the \fchic values after polarization weighting is taken as the systematic uncertainty.

Any potential $B$-meson decay contamination of the prompt \jpsi yield is checked by tightening the collision point and \chic vertex association requirement, and the results are consistent with the default selection within the statistical uncertainties. Table \ref{tab:syserrors_sources} summarizes the maximum systematic uncertainty contributions. The largest uncertainties are found at the lowest \ptjpsi range. The total systematic uncertainty corresponds to the standard deviation of the \fchic results obtained from the different systematic uncertainty sources.

\begin{table}[htb]
    \centering
    \caption{Systematic uncertainty sources and maximum individual contributions.}
    \begin{tabular}{lcc}
        Source & $1.5<y^*<4.0$ & $-5.0<y^*<-2.5$ \\\hline
       Fit parameter initial values  & $<0.050$ & $<0.058$ \\
       Shape of \chic & $<0.056$ & $<0.110$ \\
       Shape of \jpsi & $<0.002$ & $<0.002$ \\
       Mixed event normalization & $<0.03$ & $<0.010$ \\
       Kinematic input in simulation & $<0.012$ & $<0.018$ \\
       Photon efficiency & $<0.030$ & $<0.030$ \\
       Polarization of \chic & $<0.031$ & $<0.050$\\\hline 
    \end{tabular}
    \label{tab:syserrors_sources}
\end{table}

%% file: result.tex
\begin{figure}[tb]
    \nolinenumbers
    \centering
    \input{Fig2}
    \vspace{0.5cm}
    \caption{Fraction of \chic decays in the prompt \jpsi yield in \pPb and $pp$ collisions \cite{LHCb-PAPER-2011-030} as a function of \ptjpsi. The error bars show the statistical uncertainties. Boxes represent systematic uncertainties and the gray band represents the maximum uncertainties from \chic and \jpsi polarization effects.}
    \label{fig:fchic_results}
\end{figure}
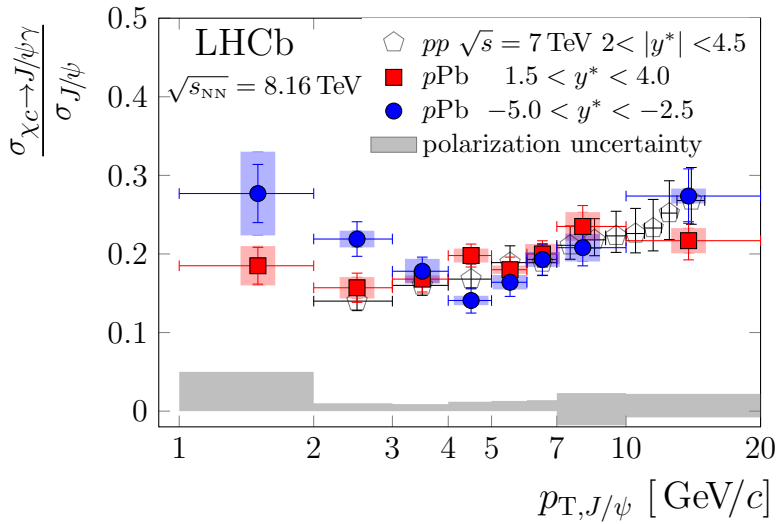

The \ptjpsi-dependent values of \fchic are shown in Fig.~\ref{fig:fchic_results}. The points are located at the mean $\ptjpsi$ of each bin determined by a fit to the prompt \jpsi differential cross-section reported by the LHCb collaboration in Ref. \cite{LHCb-PAPER-2017-014}. The new results are compared with those obtained in $pp$ collisions at \mbox{$\sqs=7$\tev~\cite{LHCb-PAPER-2011-030}}. The \fchic results obtained in \pPb and in $pp$ collisions have an overall consistency. The fraction is slightly larger in the backward rapidity region for \mbox{$\ptjpsi<3\gevc$}. 
The result is also consistent with results obtained by the HERA-B experiment in $p$C and $p$W collisions at \mbox{$\sqsnn=41.6$\gev}~\cite{HERA-B:2008rhh}. The HERA-B result covers a rapidity range \mbox{$-5.2<y^*<-2.2$} and \mbox{$\ptjpsi<2$\gevc}. A measurement of \fchic  was performed by the PHENIX collaboration in $d$Au collisions at $\sqsnn=200$\gev~\cite{PHENIX:2013pmn}, covering the mid-rapidity range\footnote{Center-of-mass and laboratory rapidities are the same in the PHENIX measurement.} $|y^*|<0.5$ and integrated over \ptjpsi, and is consistent with the $\ptjpsi<2$\gevc result presented here,  though the PHENIX measurement has large statistical uncertainties. 

\begin{table}[!htb]
    \centering
    \caption{\fchic results for wide \ptjpsi bins. The two uncertainties are statistical and systematic respectively.}
    \label{tab:fchic_result_twobins}
    \begin{tabular}{lccc}
      $\frac{\pt}{[\gevc]}$ & $\frac{<\pt>}{[\gevc]}$ & $1.5<y^*<4.0$ & $-5.0<y^*<-2.5$ \\\hline
        1--3 & 1.9 & 0.164 $\pm$ 0.016 $\pm$ 0.015 & 0.253 $\pm$ 0.019 $\pm$ 0.022 \\
       2--20 & 3.9 & 0.179 $\pm$ 0.008 $\pm$ 0.006 & 0.210 $\pm$ 0.013 $\pm$ 0.007 \\
       3--20 & 5.0 & 0.187 $\pm$ 0.008 $\pm$ 0.006 & 0.170 $\pm$ 0.013 $\pm$ 0.008 \\
       1--20 & 3.2 & 0.174 $\pm$ 0.009 $\pm$ 0.010 & 0.222 $\pm$ 0.013 $\pm$ 0.010\\\hline
    \end{tabular}
\end{table}

The difference between the backward and forward rapidity results at low \ptjpsi is further investigated by redoing the measurements using wide \ptjpsi bins to reduce statistical and systematic uncertainties. Table~\ref{tab:fchic_result_twobins} indicates that the fraction \fchic at \mbox{$\ptjpsi<3\gevc$} is \mbox{2.4\,$\sigma$} larger at backward than at forward rapidity. This apparent increase can be attributed to the suppression of \psitwos and other feeddown sources to the prompt \jpsi yield $\sigma_{\jpsi}$ in Eq.(\ref{eq:f_chic}). The fraction of \psitwos contribution to the prompt \jpsi yield is \mbox{$\frac{\sigma_{\psitwos}}{\sigma_{\jpsi}}\BR(\psitwos\to\jpsi+X)=0.080\pm 0.011$} 
according to the cross-section measured by the \lhcb collaboration in $pp$ collisions at \mbox{\sqs=7\tev} \cite{LHCb-PAPER-2011-003,LHCb-PAPER-2011-045} and the branching ratio from Ref. \cite{PDG2022}. The \mbox{$\sigma_{\psitwos}/\sigma_{\jpsi}$} ratio is reduced between \pPb collisions at backward rapidity when compared to $pp$ results \mbox{\cite{PHENIX:2022nrm,ALICE:2014cgk,LHCb-PAPER-2015-058,LHCb-PAPER-2023-024}}, consistent with \fchic being larger at backward rapidity. 

The \fchic result in \pPb collisions suggests that neither \jpsi mesons, produced directly in the collisions, nor \chic states are dissociated in the medium formed in \pPb collisions, consistent with the similar nuclear modification factor measurement of \jpsi and \Dz yields in \pPb collisions \cite{LHCb-PAPER-2017-015}. 
The non-dissociation of the \chic states leads to a constraint on the free energy (or temperature) of the system to be no larger than 180\mev in \pPb collisions, based on the smallest binding energy among the \chic states. This maximum temperature is close to the estimated freeze-out temperature in \pPb collisions, which ranges between 155--160\mev \cite{HotQCD:2018pds,Sharma:2018jqf,Flor:2021olm}. The result presented here is integrated over collision centrality. Future studies selecting central events could be more sensitive to a short-lived hot medium in \pPb collisions.

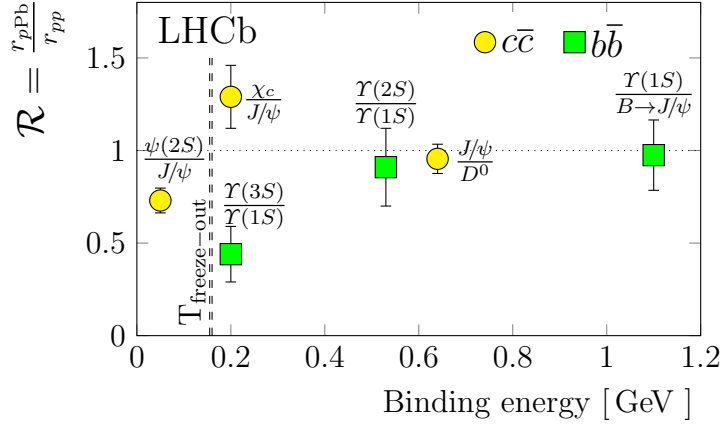
\begin{figure}[tb]
    \nolinenumbers
    \centering
    \input{Fig3}
    \vspace{0.5cm}
    \caption{Double ratio between quarkonium states vs.~binding energy along with the estimated freeze-out temperature in \pPb collisions \cite{HotQCD:2018pds,Sharma:2018jqf,Flor:2021olm}. The quarkonium state ratio $r$ is indicated for each point and described in Table \ref{tab:R_binding_energy}.}
    \label{fig:R_binding_energy}
\end{figure}

The most common method to quantify quarkonium dissociation in medium is the use of the quarkonium state yield ratios $r$ relative to their corresponding ground state \jpsi for \ccbar and \Y1S for \bbbar or to an open heavy flavor meson (\Dz and \Bp for \ccbar and \bbbar, respectively). 
The double ratio $\mathcal{R}=r_{p\rm{Pb}}/r_{pp}$ compares the $r$ values between $pp$ and \pPb collisions. 
It is obtained for \chic states from the $2<\ptjpsi<20$\gevc integrated result presented in Table \ref{tab:fchic_result_twobins} and a weighted average of the $pp$ results presented in Ref. \cite{LHCb-PAPER-2011-030}. 
The double ratio $\mathcal{R}(\chic)$ is \mbox{1.10 $\pm$ 0.12} at forward rapidity and \mbox{$1.29\pm0.17$} at backward rapidity. 
Figure \ref{fig:R_binding_energy} shows the binding-energy dependence of $\mathcal{R}$ for all quarkonium states measured by the \lhcb experiment in \pPb collisions at backward rapidity, where \pPb collisions achieve the highest particle multiplicities. The only dissociated ($\mathcal{R}<1$) quarkonium state with binding energy above the freeze-out temperature is the \Y3S. With a similar binding energy and size as the \chic states, according to non-relativistic potential theory \cite{Satz:2005hx}, the \Y3S resonance is 2.9 times heavier and may travel the medium slower than the \chic states favoring its dissociation by its interaction with comoving particles \cite{Ferreiro:2014bia}. 

\begin{table}[htb]
    \caption{Description of the double ratio measurements shown in Figure \ref{fig:R_binding_energy}.}
    \centering
    \begingroup
    \renewcommand{\arraystretch}{1.5} 
    \begin{footnotesize}
    \begin{tabular}{lcccc}
        ratio & reference & $y^*$ & \sqsnn & \pt \\\hline
        $\frac{\psitwos}{\jpsi}$ & \cite{LHCb-PAPER-2023-024} & [-5.0,-2.5] & 8.16\tev & $<14$\gevc\\
        $\frac{\chic}{\jpsi}$ & this Letter & [-5.0,-2.5] & 8.16\tev & $2<\ptjpsi<20$\gevc \\
        $\frac{\jpsi}{\Dz}$ & \cite{LHCb-PAPER-2017-015} & [-4.0,-2.5] & 5\tev & $<10$\gevc\\
        $\frac{\Y3S,\Y2S}{\Y1S}$ & \cite{LHCb-PAPER-2018-035} & [-4.5,-2.5] & 8.16\tev & $<25$\gevc\\
        $\frac{\Y1S}{B\to\jpsi}$ & \cite{LHCb-PAPER-2018-035} & [-4.5,-2.5] & 8.16\tev & $<25$\gevc\\\hline
    \end{tabular}
    \end{footnotesize}
    \endgroup
    \label{tab:R_binding_energy}
\end{table}

In summary, this Letter presents the first \lhc measurement of the fraction of \chic decays in the prompt \jpsi yield in heavy ion collisions. The ratio measured in \pPb collisions is consistent with no dissociation of \chic states, indicating that the average free energy available in these collisions is not able to inhibit the formation of quarkonium states with binding energy equal or larger than 180\mev. Such an energy is only 20--25\mev larger than the expected freeze-out temperature estimated in these systems.

%% file: Fig2.tex
\begin{tikzpicture}
   \begin{semilogxaxis}[
        name=Fchic,
        width=9.5cm,
        height=7cm,
        legend columns=1,
	xmin=0.9, xmax=20,
        ymin=-0.02, ymax=0.5,
        xlabel={\large \hspace{5cm} \ptjpsi $\left[\gevc\right]$},
        xtick={1,2,3,4,5,7,10,20},
        xticklabels={1,2,3,4,5,7,10,20},
        ylabel={\Large \hspace{3.5cm}$\frac{\sigma_{\chic\to\jpsi\gamma}}{\sigma_{\jpsi}}$},
        legend style={at={(1.0,1.0)},anchor=north east, draw=none, font=\footnotesize,
        legend cell align={left}}]

        \addplot[only marks, black, mark options={mark=pentagon, scale=2, opacity=0.5},
                    error bars/.cd, x dir=both, x explicit, y dir=both, y explicit] table [x=pt, y=F, x error=wpt, y error plus=ep, y error minus=em] {chic_ratio_pp7TeV.dat};
        \addlegendentry{$pp$ $\sqrt{s}=7\tev$ 2$<|y^*|<$4.5};
        
        \addplot[only marks, red, mark options={mark=square*, fill=red,draw=black, scale=1.5},
                    error bars/.cd, x dir=both, x explicit, y dir=both, y explicit] table  [x=pt, y=FpPb, x error minus=wptminus, x error plus=wptplus, y error=FpPb_err] {Fchic_results.dat};
        \addlegendentry{\pPb~~ ~$1.5<y^*<4.0$};
        \addplot[forget plot, only marks, red, mark options={mark=*, scale=0}, opacity=0.3,
                    error bars/.cd, x dir=both, x explicit, y dir=both, y explicit,
                    error bar style={line width=13pt}] table  [x=pt, y=FpPb, y error=FpPbsys] {Fchic_results.dat};
        
        \addplot[only marks, blue, mark options={mark=*, fill=blue, draw=black, scale=1.5},
                    error bars/.cd, x dir=both, x explicit, y dir=both, y explicit] table  [x=pt, y=FPbp, x error minus=wptminus, x error plus=wptplus, y error=FPbp_err] {Fchic_results.dat};
        \addlegendentry{\pPb~ $-5.0<y^*<-2.5$};
        \addplot[forget plot, only marks, blue, mark options={mark=*, scale=0}, opacity=0.3,
                    error bars/.cd, x dir=both, x explicit, y dir=both, y explicit,
                    error bar style={line width=13pt}] table  [x=pt, y=FPbp, y error=FPbpsys] {Fchic_results.dat};

        \addplot [name path=upper,draw=none,forget plot] coordinates {
            (1,0.05)
            (2,0.05)
            (2,0.01)
            (3,0.01)
            (3,0.009)
            (4,0.009) 
            (4,0.012)
            (5,0.012)
            (5,0.013)
            (6,0.013)
            (6,0.014)
            (7,0.014)
            (7,0.023)
            (10,0.023)
            (10,0.022)
            (20,0.022)
        };
        \addplot [name path=lower,draw=none,forget plot] coordinates {
            (1,0)
            (2,0)
            (2,0)
            (3,0)
            (3,0)
            (4,0)
            (4,0)
            (5,0)
            (5,0)
            (6,0)
            (6,0)
            (7,0)
            (7,-0.018)
            (10,-0.018)
            (10,-0.008)
            (20,-0.008)
        };
        \addplot +[fill=gray, opacity=0.5] fill between[of=lower and upper];
        \addlegendentry{polarization uncertainty};
        \node at (axis cs: 1.4, 0.47) {\large \lhcb};
        \node at (axis cs: 1.55, 0.41) {\footnotesize $\sqsnn=8.16\tev$};
  \end{semilogxaxis}
\end{tikzpicture}
\vspace{-0.7cm}

%% file: Fig3.tex
 \begin{tikzpicture}
 \begin{axis}[
    width=9cm,
    height=6cm,
    ymax = 1.8,
    ymin = 0.0,
    xmin = 0,
    xmax = 1.2,
    xlabel={\hspace{3cm} Binding energy [\gev]},
    ylabel={\large \hspace{2.5cm} $\mathcal{R}=\frac{r_{p\rm{Pb}}}{r_{pp}}$},
    legend style={at={(axis cs:0.7,1.75)}, draw=none, legend cell align={left}, anchor=north west, font=\large,},
    legend columns=3]
    \addplot[only marks, black, mark options={mark=*, fill=yellow,draw=black, scale=2},
            error bars/.cd, x dir=both, x explicit, y dir=both, y explicit] coordinates{
            (0.05, 0.73) +- (0, 0.067)
            (0.2, 1.29) +- (0, 0.17)
            (0.64, 0.955) +- (0, 0.079)
            };
    \addplot[only marks, black, mark options={mark=square*, fill=green,draw=black, scale=2},
            error bars/.cd, x dir=both, x explicit, y dir=both, y explicit] coordinates{
            (0.20, 0.44) +- (0., 0.15)
            (0.53, 0.91) +- (0.0, 0.21)
            (1.10, 0.975) +- (0.0, 0.19)
            };

    \addplot [draw=black, dashed] coordinates {(0.155,0) (0.155,1.5) };
    \addplot [draw=black, dashed] coordinates {(0.160,0) (0.160,1.5) };
    \addplot [draw=black, dotted] coordinates {(0,1) (1.2,1)};
    \node at (axis cs: 0.08, 0.95) {$\frac{\psitwos}{\jpsi}$};
    \node at (axis cs: 0.27, 1.25) {$\frac{\chic}{\jpsi}$};
    \node at (axis cs: 0.25, 0.7) {$\frac{\Y3S}{\Y1S}$};
    \node at (axis cs: 0.53, 1.25) {$\frac{\Y2S}{\Y1S}$};
    \node at (axis cs: 0.72, 0.95) {$\frac{\jpsi}{\Dz}$};
    \node at (axis cs: 1.1, 1.3) {$\frac{\Y1S}{B\to\jpsi}$};
    \node at (axis cs: 0.15, 1.65) {\large \lhcb};
    \node [rotate=90] at (axis cs: 0.12,0.4) {\normalsize T$_{\rm freeze-out}$};
  \legend{\ccbar~~~, \bbbar};
  \end{axis}
  \end{tikzpicture}
  \vspace{-0.7 cm}

%% file: acknowledgements.tex
\section*{Acknowledgements}
%
%
\noindent We express our gratitude to our colleagues in the CERN
accelerator departments for the excellent performance of the LHC. We
thank the technical and administrative staff at the LHCb
institutes.
We acknowledge support from CERN and from the national agencies:
CAPES, CNPq, FAPERJ and FINEP (Brazil); 
MOST and NSFC (China); 
CNRS/IN2P3 (France); 
BMBF, DFG and MPG (Germany); 
INFN (Italy); 
NWO (Netherlands); 
MNiSW and NCN (Poland); 
MCID/IFA (Romania); 
MICINN (Spain); 
SNSF and SER (Switzerland); 
NASU (Ukraine); 
STFC (United Kingdom); 
DOE NP and NSF (USA).
We acknowledge the computing resources that are provided by CERN, IN2P3
(France), KIT and DESY (Germany), INFN (Italy), SURF (Netherlands),
PIC (Spain), GridPP (United Kingdom), 
CSCS (Switzerland), IFIN-HH (Romania), CBPF (Brazil),
and Polish WLCG (Poland).
We are indebted to the communities behind the multiple open-source
software packages on which we depend.
Individual groups or members have received support from
ARC and ARDC (Australia);
Key Research Program of Frontier Sciences of CAS, CAS PIFI, CAS CCEPP, 
Fundamental Research Funds for the Central Universities, 
and Sci. \& Tech. Program of Guangzhou (China);
Minciencias (Colombia);
EPLANET, Marie Sk\l{}odowska-Curie Actions, ERC and NextGenerationEU (European Union);
A*MIDEX, ANR, IPhU and Labex P2IO, and R\'{e}gion Auvergne-Rh\^{o}ne-Alpes (France);
AvH Foundation (Germany);
ICSC (Italy); 
GVA, XuntaGal, GENCAT, Inditex, InTalent and Prog.~Atracci\'on Talento, CM (Spain);
SRC (Sweden);
the Leverhulme Trust, the Royal Society
 and UKRI (United Kingdom).

%% file: main.bbl
\ifx\mcitethebibliography\mciteundefinedmacro
\PackageError{LHCb.bst}{mciteplus.sty has not been loaded}
{This bibstyle requires the use of the mciteplus package.}\fi
\providecommand{\href}[2]{#2}

%% file: Authorship_LHCb-PAPER-2023-028.tex
\centerline
{\large\bf LHCb collaboration}
\begin
{flushleft}
\small
R.~Aaij$^{35}$\lhcborcid{0000-0003-0533-1952},
A.S.W.~Abdelmotteleb$^{54}$\lhcborcid{0000-0001-7905-0542},
C.~Abellan~Beteta$^{48}$,
F.~Abudin{\'e}n$^{54}$\lhcborcid{0000-0002-6737-3528},
T.~Ackernley$^{58}$\lhcborcid{0000-0002-5951-3498},
B.~Adeva$^{44}$\lhcborcid{0000-0001-9756-3712},
M.~Adinolfi$^{52}$\lhcborcid{0000-0002-1326-1264},
P.~Adlarson$^{78}$\lhcborcid{0000-0001-6280-3851},
C.~Agapopoulou$^{46}$\lhcborcid{0000-0002-2368-0147},
C.A.~Aidala$^{79}$\lhcborcid{0000-0001-9540-4988}
% ...

{\footnotesize \it

% ...
$^{35}$Nikhef National Institute for Subatomic Physics, Amsterdam, Netherlands\\
% ...
$^{44}$Instituto Galego de F{\'\i}sica de Altas Enerx{\'\i}as (IGFAE), Universidade de Santiago de Compostela, Santiago de Compostela, Spain\\
% ...
$^{46}$European Organization for Nuclear Research (CERN), Geneva, Switzerland\\
% ...
$^{48}$Physik-Institut, Universit{\"a}t Z{\"u}rich, Z{\"u}rich, Switzerland\\
% ...
$^{52}$H.H. Wills Physics Laboratory, University of Bristol, Bristol, United Kingdom\\
% ...
$^{54}$Department of Physics, University of Warwick, Coventry, United Kingdom\\
% ...
$^{57}$School of Physics and Astronomy, University of Glasgow, Glasgow, United Kingdom\\
$^{58}$Oliver Lodge Laboratory, University of Liverpool, Liverpool, United Kingdom\\
% ...
$^{66}$Syracuse University, Syracuse, NY, United States\\
% ...
$^{78}$Department of Physics and Astronomy, Uppsala University, Uppsala, Sweden, associated to $^{57}$\\
$^{79}$University of Michigan, Ann Arbor, MI, United States, associated to $^{66}$\\
% ...
}
\end{flushleft}

%% file: main.bbl
\begin{mcitethebibliography}{10}
\mciteSetBstSublistMode{n}
\mciteSetBstMaxWidthForm{subitem}{\alph{mcitesubitemcount})}
\mciteSetBstSublistLabelBeginEnd{\mcitemaxwidthsubitemform\space}
{\relax}{\relax}

\bibitem{Annala:2019puf}
E.~Annala {\em et~al.}, \ifthenelse{\boolean{articletitles}}{\emph{{Evidence
  for quark-matter cores in massive neutron stars}},
  }{}\href{https://doi.org/10.1038/s41567-020-0914-9}{Nature Phys.\
  \textbf{16} (2020) 907},
  \href{http://arxiv.org/abs/1903.09121}{{\normalfont\ttfamily
  arXiv:1903.09121}}\relax
\mciteBstWouldAddEndPuncttrue
\mciteSetBstMidEndSepPunct{\mcitedefaultmidpunct}
{\mcitedefaultendpunct}{\mcitedefaultseppunct}\relax
\EndOfBibitem
\bibitem{PHENIX:2018lia}
PHENIX collaboration, C.~Aidala {\em et~al.},
  \ifthenelse{\boolean{articletitles}}{\emph{{Creation of
  quark\textendash{}gluon plasma droplets with three distinct geometries}},
  }{}\href{https://doi.org/10.1038/s41567-018-0360-0}{Nature Phys.\
  \textbf{15} (2019) 214},
  \href{http://arxiv.org/abs/1805.02973}{{\normalfont\ttfamily
  arXiv:1805.02973}}\relax
\mciteBstWouldAddEndPuncttrue
\mciteSetBstMidEndSepPunct{\mcitedefaultmidpunct}
{\mcitedefaultendpunct}{\mcitedefaultseppunct}\relax
\EndOfBibitem
\bibitem{Satz:2005hx}
H.~Satz, \ifthenelse{\boolean{articletitles}}{\emph{{Colour deconfinement and
  quarkonium binding}},
  }{}\href{https://doi.org/10.1088/0954-3899/32/3/R01}{J.\ Phys.\  \textbf{G32}
  (2006) R25}\relax
\mciteBstWouldAddEndPuncttrue
\mciteSetBstMidEndSepPunct{\mcitedefaultmidpunct}
{\mcitedefaultendpunct}{\mcitedefaultseppunct}\relax
\EndOfBibitem
\bibitem{HotQCD:2018pds}
HotQCD collaboration, A.~Bazavov {\em et~al.},
  \ifthenelse{\boolean{articletitles}}{\emph{{Chiral crossover in QCD at zero
  and non-zero chemical potentials}},
  }{}\href{https://doi.org/10.1016/j.physletb.2019.05.013}{Phys.\ Lett.\
  \textbf{B795} (2019) 15},
  \href{http://arxiv.org/abs/1812.08235}{{\normalfont\ttfamily
  arXiv:1812.08235}}\relax
\mciteBstWouldAddEndPuncttrue
\mciteSetBstMidEndSepPunct{\mcitedefaultmidpunct}
{\mcitedefaultendpunct}{\mcitedefaultseppunct}\relax
\EndOfBibitem
\bibitem{Sharma:2018jqf}
N.~Sharma, J.~Cleymans, B.~Hippolyte, and M.~Paradza,
  \ifthenelse{\boolean{articletitles}}{\emph{{A Comparison of p-p, p-Pb, Pb-Pb
  collisions in the thermal model: multiplicity dependence of thermal
  parameters}}, }{}\href{https://doi.org/10.1103/PhysRevC.99.044914}{Phys.\
  Rev.\  \textbf{C99} (2019) 044914},
  \href{http://arxiv.org/abs/1811.00399}{{\normalfont\ttfamily
  arXiv:1811.00399}}\relax
\mciteBstWouldAddEndPuncttrue
\mciteSetBstMidEndSepPunct{\mcitedefaultmidpunct}
{\mcitedefaultendpunct}{\mcitedefaultseppunct}\relax
\EndOfBibitem
\bibitem{Flor:2021olm}
F.~A. Flor, G.~Olinger, and R.~Bellwied,
  \ifthenelse{\boolean{articletitles}}{\emph{{System size and flavour
  dependence of chemical freeze-out temperatures in ALICE data from pp, pPb and
  PbPb collisions at LHC energies}},
  }{}\href{https://doi.org/10.1016/j.physletb.2022.137473}{Phys.\ Lett.\
  \textbf{B834} (2022) 137473},
  \href{http://arxiv.org/abs/2109.09843}{{\normalfont\ttfamily
  arXiv:2109.09843}}\relax
\mciteBstWouldAddEndPuncttrue
\mciteSetBstMidEndSepPunct{\mcitedefaultmidpunct}
{\mcitedefaultendpunct}{\mcitedefaultseppunct}\relax
\EndOfBibitem
\bibitem{Ferreiro:2014bia}
E.~G. Ferreiro, \ifthenelse{\boolean{articletitles}}{\emph{{Excited charmonium
  suppression in proton–nucleus collisions as a consequence of comovers}},
  }{}\href{https://doi.org/10.1016/j.physletb.2015.07.066}{Phys.\ Lett.\
  \textbf{B749} (2015) 98},
  \href{http://arxiv.org/abs/1411.0549}{{\normalfont\ttfamily
  arXiv:1411.0549}}\relax
\mciteBstWouldAddEndPuncttrue
\mciteSetBstMidEndSepPunct{\mcitedefaultmidpunct}
{\mcitedefaultendpunct}{\mcitedefaultseppunct}\relax
\EndOfBibitem
\bibitem{LHCb-PAPER-2017-015}
LHCb collaboration, R.~Aaij {\em et~al.},
  \ifthenelse{\boolean{articletitles}}{\emph{{Study of prompt \Dz meson
  production in \proton{}Pb collisions at $\sqsnn = $5\tev}},
  }{}\href{https://doi.org/10.1007/JHEP10(2017)090}{JHEP \textbf{10} (2017)
  090}, \href{http://arxiv.org/abs/1707.02750}{{\normalfont\ttfamily
  arXiv:1707.02750}}\relax
\mciteBstWouldAddEndPuncttrue
\mciteSetBstMidEndSepPunct{\mcitedefaultmidpunct}
{\mcitedefaultendpunct}{\mcitedefaultseppunct}\relax
\EndOfBibitem
\bibitem{PHENIX:2022nrm}
PHENIX collaboration, U.~A. Acharya {\em et~al.},
  \ifthenelse{\boolean{articletitles}}{\emph{{Measurement of $\psi$(2S) nuclear
  modification at backward and forward rapidity in $p+p$, $p$+Al, and $p$+Au
  collisions at $\sqrt{s_{NN}}$=200 GeV}},
  }{}\href{https://doi.org/10.1103/PhysRevC.105.064912}{Phys.\ Rev.\
  \textbf{C105} (2022) 064912},
  \href{http://arxiv.org/abs/2202.03863}{{\normalfont\ttfamily
  arXiv:2202.03863}}\relax
\mciteBstWouldAddEndPuncttrue
\mciteSetBstMidEndSepPunct{\mcitedefaultmidpunct}
{\mcitedefaultendpunct}{\mcitedefaultseppunct}\relax
\EndOfBibitem
\bibitem{LHCb-PAPER-2015-058}
LHCb collaboration, R.~Aaij {\em et~al.},
  \ifthenelse{\boolean{articletitles}}{\emph{{Study of \psitwos production
  cross-sections and cold nuclear matter effects in \proton{}Pb collisions at
  $\sqsnn = $5\tev}}, }{}\href{https://doi.org/10.1007/JHEP03(2016)133}{JHEP
  \textbf{03} (2016) 133},
  \href{http://arxiv.org/abs/1601.07878}{{\normalfont\ttfamily
  arXiv:1601.07878}}\relax
\mciteBstWouldAddEndPuncttrue
\mciteSetBstMidEndSepPunct{\mcitedefaultmidpunct}
{\mcitedefaultendpunct}{\mcitedefaultseppunct}\relax
\EndOfBibitem
\bibitem{ALICE:2014cgk}
ALICE collaboration, B.~B. Abelev {\em et~al.},
  \ifthenelse{\boolean{articletitles}}{\emph{{Suppression of $\psi$(2S)
  production in p-Pb collisions at $\sqrt{s_{\rm NN}}$ = 5.02 TeV}},
  }{}\href{https://doi.org/10.1007/JHEP12(2014)073}{JHEP \textbf{12} (2014)
  073}, \href{http://arxiv.org/abs/1405.3796}{{\normalfont\ttfamily
  arXiv:1405.3796}}\relax
\mciteBstWouldAddEndPuncttrue
\mciteSetBstMidEndSepPunct{\mcitedefaultmidpunct}
{\mcitedefaultendpunct}{\mcitedefaultseppunct}\relax
\EndOfBibitem
\bibitem{LHCb-PAPER-2023-024}
LHCb collaboration, R.~Aaij {\em et~al.},
  \ifthenelse{\boolean{articletitles}}{\emph{{Prompt and nonprompt $\psi(2S)$
  production in $p$Pb collisions at $\sqrt{s_{NN}} = 8.16$ TeV}}, }{}
  {LHCb-PAPER-2023-024}, {To be submitted to JHEP}\relax
\mciteBstWouldAddEndPuncttrue
\mciteSetBstMidEndSepPunct{\mcitedefaultmidpunct}
{\mcitedefaultendpunct}{\mcitedefaultseppunct}\relax
\EndOfBibitem
\bibitem{PDG2022}
Particle Data Group, R.~L. Workman {\em et~al.},
  \ifthenelse{\boolean{articletitles}}{\emph{{Review of Particle Physics}},
  }{}\href{https://doi.org/10.1093/ptep/ptac097}{PTEP \textbf{2022} (2022)
  083C01}\relax
\mciteBstWouldAddEndPuncttrue
\mciteSetBstMidEndSepPunct{\mcitedefaultmidpunct}
{\mcitedefaultendpunct}{\mcitedefaultseppunct}\relax
\EndOfBibitem
\bibitem{HERA-B:2008rhh}
HERA-B collaboration, I.~Abt {\em et~al.},
  \ifthenelse{\boolean{articletitles}}{\emph{{Production of the charmonium
  states $\chi_{c1}$ and $\chi_{c2}$ in proton nucleus interactions at
  $\sqrt{s}$ = 41.6-GeV}},
  }{}\href{https://doi.org/10.1103/PhysRevD.79.012001}{Phys.\ Rev.\
  \textbf{D79} (2009) 012001},
  \href{http://arxiv.org/abs/0807.2167}{{\normalfont\ttfamily
  arXiv:0807.2167}}\relax
\mciteBstWouldAddEndPuncttrue
\mciteSetBstMidEndSepPunct{\mcitedefaultmidpunct}
{\mcitedefaultendpunct}{\mcitedefaultseppunct}\relax
\EndOfBibitem
\bibitem{PHENIX:2013pmn}
PHENIX collaboration, A.~Adare {\em et~al.},
  \ifthenelse{\boolean{articletitles}}{\emph{{Nuclear modification of
  $\psi^{\prime}$, $\chi_c$, and J/$\ensuremath{\psi}$ production in d+Au
  collisions at $\sqrt{s_{NN}}$=200 GeV}},
  }{}\href{https://doi.org/10.1103/PhysRevLett.111.202301}{Phys.\ Rev.\ Lett.\
  \textbf{111} (2013) 202301},
  \href{http://arxiv.org/abs/1305.5516}{{\normalfont\ttfamily
  arXiv:1305.5516}}\relax
\mciteBstWouldAddEndPuncttrue
\mciteSetBstMidEndSepPunct{\mcitedefaultmidpunct}
{\mcitedefaultendpunct}{\mcitedefaultseppunct}\relax
\EndOfBibitem
\bibitem{LHCb-PAPER-2020-048}
LHCb collaboration, R.~Aaij {\em et~al.},
  \ifthenelse{\boolean{articletitles}}{\emph{{Measurement of
  prompt-cross-section ratio $\sigma(\chi_{c2})/\sigma(\chi_{c1})$ in $p$Pb
  collisions at $\sqrt{s_{NN}}$ = 8.16 TeV}},
  }{}\href{https://doi.org/10.1103/PhysRevC.103.064905}{Phys.\ Rev.\
  \textbf{C103} (2021) 064905},
  \href{http://arxiv.org/abs/2103.07349}{{\normalfont\ttfamily
  arXiv:2103.07349}}\relax
\mciteBstWouldAddEndPuncttrue
\mciteSetBstMidEndSepPunct{\mcitedefaultmidpunct}
{\mcitedefaultendpunct}{\mcitedefaultseppunct}\relax
\EndOfBibitem
\bibitem{LHCb-DP-2008-001}
LHCb collaboration, A.~A. Alves~Jr.\ {\em et~al.},
  \ifthenelse{\boolean{articletitles}}{\emph{{The \lhcb detector at the LHC}},
  }{}\href{https://doi.org/10.1088/1748-0221/3/08/S08005}{JINST \textbf{3}
  (2008) S08005}\relax
\mciteBstWouldAddEndPuncttrue
\mciteSetBstMidEndSepPunct{\mcitedefaultmidpunct}
{\mcitedefaultendpunct}{\mcitedefaultseppunct}\relax
\EndOfBibitem
\bibitem{LHCb-DP-2014-002}
LHCb collaboration, R.~Aaij {\em et~al.},
  \ifthenelse{\boolean{articletitles}}{\emph{{LHCb detector performance}},
  }{}\href{https://doi.org/10.1142/S0217751X15300227}{Int.\ J.\ Mod.\ Phys.\
  \textbf{A30} (2015) 1530022},
  \href{http://arxiv.org/abs/1412.6352}{{\normalfont\ttfamily
  arXiv:1412.6352}}\relax
\mciteBstWouldAddEndPuncttrue
\mciteSetBstMidEndSepPunct{\mcitedefaultmidpunct}
{\mcitedefaultendpunct}{\mcitedefaultseppunct}\relax
\EndOfBibitem
\bibitem{LHCb-PROC-2014-017}
P.~Perret, \ifthenelse{\boolean{articletitles}}{\emph{{First Years of Running
  for the LHCb Calorimeter system}}, }{}
  \href{http://cdsweb.cern.ch/search?p=LHCb-PROC-2014-017&f=reportnumber&action_search=Search&c=LHCb+Conference+Proceedings}
  {LHCb-PROC-2014-017}, 2014\relax
\mciteBstWouldAddEndPuncttrue
\mciteSetBstMidEndSepPunct{\mcitedefaultmidpunct}
{\mcitedefaultendpunct}{\mcitedefaultseppunct}\relax
\EndOfBibitem
\bibitem{Sjostrand:2007gs}
T.~Sj\"{o}strand, S.~Mrenna, and P.~Skands,
  \ifthenelse{\boolean{articletitles}}{\emph{{A brief introduction to PYTHIA
  8.1}}, }{}\href{https://doi.org/10.1016/j.cpc.2008.01.036}{Comput.\ Phys.\
  Commun.\  \textbf{178} (2008) 852},
  \href{http://arxiv.org/abs/0710.3820}{{\normalfont\ttfamily
  arXiv:0710.3820}}\relax
\mciteBstWouldAddEndPuncttrue
\mciteSetBstMidEndSepPunct{\mcitedefaultmidpunct}
{\mcitedefaultendpunct}{\mcitedefaultseppunct}\relax
\EndOfBibitem
\bibitem{Pierog:2013ria}
T.~Pierog {\em et~al.}, \ifthenelse{\boolean{articletitles}}{\emph{{EPOS LHC:
  Test of collective hadronization with data measured at the CERN Large Hadron
  Collider}}, }{}\href{https://doi.org/10.1103/PhysRevC.92.034906}{Phys.\ Rev.\
   \textbf{C92} (2015) 034906},
  \href{http://arxiv.org/abs/1306.0121}{{\normalfont\ttfamily
  arXiv:1306.0121}}\relax
\mciteBstWouldAddEndPuncttrue
\mciteSetBstMidEndSepPunct{\mcitedefaultmidpunct}
{\mcitedefaultendpunct}{\mcitedefaultseppunct}\relax
\EndOfBibitem
\bibitem{Lange:2001uf}
D.~J. Lange, \ifthenelse{\boolean{articletitles}}{\emph{{The EvtGen particle
  decay simulation package}},
  }{}\href{https://doi.org/10.1016/S0168-9002(01)00089-4}{Nucl.\ Instrum.\
  Meth.\  \textbf{A462} (2001) 152}\relax
\mciteBstWouldAddEndPuncttrue
\mciteSetBstMidEndSepPunct{\mcitedefaultmidpunct}
{\mcitedefaultendpunct}{\mcitedefaultseppunct}\relax
\EndOfBibitem
\bibitem{Golonka:2005pn}
P.~Golonka and Z.~Was, \ifthenelse{\boolean{articletitles}}{\emph{{PHOTOS Monte
  Carlo: A precision tool for QED corrections in $Z$ and $W$ decays}},
  }{}\href{https://doi.org/10.1140/epjc/s2005-02396-4}{Eur.\ Phys.\ J.\
  \textbf{C45} (2006) 97},
  \href{http://arxiv.org/abs/hep-ph/0506026}{{\normalfont\ttfamily
  arXiv:hep-ph/0506026}}\relax
\mciteBstWouldAddEndPuncttrue
\mciteSetBstMidEndSepPunct{\mcitedefaultmidpunct}
{\mcitedefaultendpunct}{\mcitedefaultseppunct}\relax
\EndOfBibitem
\bibitem{LHCb-PROC-2011-006}
M.~Clemencic {\em et~al.}, \ifthenelse{\boolean{articletitles}}{\emph{{The
  \lhcb simulation application, Gauss: Design, evolution and experience}},
  }{}\href{https://doi.org/10.1088/1742-6596/331/3/032023}{J.\ Phys.\ Conf.\
  Ser.\  \textbf{331} (2011) 032023}\relax
\mciteBstWouldAddEndPuncttrue
\mciteSetBstMidEndSepPunct{\mcitedefaultmidpunct}
{\mcitedefaultendpunct}{\mcitedefaultseppunct}\relax
\EndOfBibitem
\bibitem{Skwarnicki:1986xj}
T.~Skwarnicki, {\em {A study of the radiative cascade transitions between the
  Upsilon-prime and Upsilon resonances}}, PhD thesis, Institute of Nuclear
  Physics, Krakow, 1986,
  {\href{http://inspirehep.net/record/230779/}{DESY-F31-86-02}}\relax
\mciteBstWouldAddEndPuncttrue
\mciteSetBstMidEndSepPunct{\mcitedefaultmidpunct}
{\mcitedefaultendpunct}{\mcitedefaultseppunct}\relax
\EndOfBibitem
\bibitem{Eskola:2021nhw}
K.~J. Eskola, P.~Paakkinen, H.~Paukkunen, and C.~A. Salgado,
  \ifthenelse{\boolean{articletitles}}{\emph{{EPPS21: a global QCD analysis of
  nuclear PDFs}},
  }{}\href{https://doi.org/10.1140/epjc/s10052-022-10359-0}{Eur.\ Phys.\ J.\
  \textbf{C82} (2022) 413},
  \href{http://arxiv.org/abs/2112.12462}{{\normalfont\ttfamily
  arXiv:2112.12462}}\relax
\mciteBstWouldAddEndPuncttrue
\mciteSetBstMidEndSepPunct{\mcitedefaultmidpunct}
{\mcitedefaultendpunct}{\mcitedefaultseppunct}\relax
\EndOfBibitem
\bibitem{LHCb-PAPER-2011-030}
LHCb collaboration, R.~Aaij {\em et~al.},
  \ifthenelse{\boolean{articletitles}}{\emph{{Measurement of the ratio of
  prompt $\chi_{c}$ to \jpsi production in \proton\proton collisions at
  \mbox{$\sqs=$7\tev}}},
  }{}\href{https://doi.org/10.1016/j.physletb.2012.10.068}{Phys.\ Lett.\
  \textbf{B718} (2012) 431},
  \href{http://arxiv.org/abs/1204.1462}{{\normalfont\ttfamily
  arXiv:1204.1462}}\relax
\mciteBstWouldAddEndPuncttrue
\mciteSetBstMidEndSepPunct{\mcitedefaultmidpunct}
{\mcitedefaultendpunct}{\mcitedefaultseppunct}\relax
\EndOfBibitem
\bibitem{LHCb-PAPER-2013-008}
LHCb collaboration, R.~Aaij {\em et~al.},
  \ifthenelse{\boolean{articletitles}}{\emph{{Measurement of \jpsi polarization
  in \proton\proton collisions at \mbox{$\sqs=$7\tev}}},
  }{}\href{https://doi.org/10.1140/epjc/s10052-013-2631-3}{Eur.\ Phys.\ J.\
  \textbf{C73} (2013) 2631},
  \href{http://arxiv.org/abs/1307.6379}{{\normalfont\ttfamily
  arXiv:1307.6379}}\relax
\mciteBstWouldAddEndPuncttrue
\mciteSetBstMidEndSepPunct{\mcitedefaultmidpunct}
{\mcitedefaultendpunct}{\mcitedefaultseppunct}\relax
\EndOfBibitem
\bibitem{CMS:2019jas}
CMS collaboration, A.~M. Sirunyan {\em et~al.},
  \ifthenelse{\boolean{articletitles}}{\emph{{Constraints on the
  $\chi_\mathrm{c1}$ versus $\chi_\mathrm{c2}$ polarizations in proton-proton
  collisions at $\sqrt{s} =$ 8 TeV}},
  }{}\href{https://doi.org/10.1103/PhysRevLett.124.162002}{Phys.\ Rev.\ Lett.\
  \textbf{124} (2020) 162002},
  \href{http://arxiv.org/abs/1912.07706}{{\normalfont\ttfamily
  arXiv:1912.07706}}\relax
\mciteBstWouldAddEndPuncttrue
\mciteSetBstMidEndSepPunct{\mcitedefaultmidpunct}
{\mcitedefaultendpunct}{\mcitedefaultseppunct}\relax
\EndOfBibitem
\bibitem{LHCb-PAPER-2017-014}
LHCb collaboration, R.~Aaij {\em et~al.},
  \ifthenelse{\boolean{articletitles}}{\emph{{Prompt and nonprompt \jpsi\
  production and nuclear modification in \proton{}Pb collisions at $\sqsnn =
  $8.16\tev}}, }{}\href{https://doi.org/10.1016/j.physletb.2017.09.058}{Phys.\
  Lett.\  \textbf{B774} (2017) 159},
  \href{http://arxiv.org/abs/1706.07122}{{\normalfont\ttfamily
  arXiv:1706.07122}}\relax
\mciteBstWouldAddEndPuncttrue
\mciteSetBstMidEndSepPunct{\mcitedefaultmidpunct}
{\mcitedefaultendpunct}{\mcitedefaultseppunct}\relax
\EndOfBibitem
\bibitem{LHCb-PAPER-2011-003}
LHCb collaboration, R.~Aaij {\em et~al.},
  \ifthenelse{\boolean{articletitles}}{\emph{{Measurement of \jpsi production
  in \proton\proton collisions at \mbox{$\sqs=$7\tev}}},
  }{}\href{https://doi.org/10.1140/epjc/s10052-011-1645-y}{Eur.\ Phys.\ J.\
  \textbf{C71} (2011) 1645},
  \href{http://arxiv.org/abs/1103.0423}{{\normalfont\ttfamily
  arXiv:1103.0423}}\relax
\mciteBstWouldAddEndPuncttrue
\mciteSetBstMidEndSepPunct{\mcitedefaultmidpunct}
{\mcitedefaultendpunct}{\mcitedefaultseppunct}\relax
\EndOfBibitem
\bibitem{LHCb-PAPER-2011-045}
LHCb collaboration, R.~Aaij {\em et~al.},
  \ifthenelse{\boolean{articletitles}}{\emph{{Measurement of \psitwos meson
  production in \proton\proton collisions at \mbox{$\sqs=$7\tev}}},
  }{}\href{https://doi.org/10.1140/epjc/s10052-012-2100-4}{Eur.\ Phys.\ J.\
  \textbf{C72} (2012) 2100}, Erratum
  \href{https://doi.org/10.1140/epjc/s10052-012-2100-4}{ibid.\   \textbf{C80}
  (2020) 49}, \href{http://arxiv.org/abs/1204.1258}{{\normalfont\ttfamily
  arXiv:1204.1258}}\relax
\mciteBstWouldAddEndPuncttrue
\mciteSetBstMidEndSepPunct{\mcitedefaultmidpunct}
{\mcitedefaultendpunct}{\mcitedefaultseppunct}\relax
\EndOfBibitem
\bibitem{LHCb-PAPER-2018-035}
LHCb collaboration, R.~Aaij {\em et~al.},
  \ifthenelse{\boolean{articletitles}}{\emph{{Study of \Upsilonres production
  in \proton{}Pb collisions at $\sqsnn = $8.16\tev}},
  }{}\href{https://doi.org/10.1007/JHEP11(2018)194}{JHEP \textbf{11} (2018)
  194}, \href{http://arxiv.org/abs/1810.07655}{{\normalfont\ttfamily
  arXiv:1810.07655}}\relax
\mciteBstWouldAddEndPuncttrue
\mciteSetBstMidEndSepPunct{\mcitedefaultmidpunct}
{\mcitedefaultendpunct}{\mcitedefaultseppunct}\relax
\EndOfBibitem
\end{mcitethebibliography}
